%
%
%
 
\ifx\mnmacrosloaded\undefined \input mn\fi 
%
%
%
 
\newif\ifAMStwofonts 
 
\ifCUPmtplainloaded \else 
  \NewTextAlphabet{textbfit} {cmbxti10} {} 
  \NewTextAlphabet{textbfss} {cmssbx10} {} 
  \NewMathAlphabet{mathbfit} {cmbxti10} {} 
  \NewMathAlphabet{mathbfss} {cmssbx10} {} 
  \ifAMStwofonts 
    \NewSymbolFont{upmath} {eurm10} 
    \NewSymbolFont{AMSa} {msam10} 
    \NewMathSymbol{\upi}     {0}{upmath}{19} 
    \NewMathSymbol{\umu}     {0}{upmath}{16} 
    \NewMathSymbol{\upartial}{0}{upmath}{40} 
    \NewMathSymbol{\leqslant}{3}{AMSa}{36} 
    \NewMathSymbol{\geqslant}{3}{AMSa}{3E}

     \let\le=\leqslant 
      
  \else 
    \def\umu{\mu} 
    \def\upi{\pi} 
    \def\upartial{\partial} 
  \fi 
\fi

 
 
\loadboldmathnames 
 
 
 
\onecolumn        
 
%
%
%
%
\def\PBvp #1 #2{ #1, #2} 
\def\PBa #1:#2 #3 #4 {#1,#2, {A\&A} \PBvp #3 #4} 
\def\PBapj #1:#2 #3 #4 {#1,#2, {ApJ} \PBvp #3 #4} 
\def\PBasupl #1:#2 #3 #4 {#1,#2, {A\&AS} \PBvp #3 #4} 
\def\PBapjsupl #1:#2 #3 #4 {#1,#2, {ApJS} \PBvp #3 #4} 
\def\PBpasp #1:#2 #3 #4 {#1,#2, { PASP} \PBvp #3 #4} 
\def\PBpaspc #1:#2 #3 #4 {#1,#2, { PASPC } \PBvp #3 #4} 
\def\PBmn #1:#2 #3 #4 {#1,#2, {MNRAS} \PBvp #3 #4} 
\def\PBmsait #1:#2 #3 #4 {#1,#2, {Mem. Soc. Astron. It.} \PBvp #3 #4} 
\def\PBnat #1:#2 #3 #4 {#1,#2, {Nat} \PBvp #3 #4} 
\def\PBaj #1:#2 #3 #4 {#1,#2, {AJ} \PBvp #3 #4} 
\def\PBjaa #1:#2 #3 #4 {#1,#2, {JA\& A} \PBvp #3 #4} 
\def\PBaspsc #1:#2 #3 #4 {#1,#2, {Ap\&SS} \PBvp #3 #4} 
\def\PBanrev #1:#2 #3 #4 {#1,#2, {ARA\&A} \PBvp #3 #4} 
\def\PBrevmex #1:#2 #3 #4 {#1,#2, {Rev. Mex. de Astron. y Astrof.} \PBvp #3 #4} 
\def\PBscie #1:#2 #3 #4 {#1,#2, {Sci} \PBvp #3 #4} 
\def\PBesomsg #1:#2 #3 #4 {#1,#2, {The Messenger} \PBvp #3 #4} 
\def\PBrmp #1:#2 #3 #4 {#1,#2, {Rev. Mod. Phys.} \PBvp #3 #4} 
\def\PBans #1:#2 #3 #4 {#1,#2, {Ann. Rev. of Nucl. Sci.} \PBvp #3 #4} 
\def\PBphrev #1:#2 #3 #4 {#1,#2, {Phys. Rev.} \PBvp #3 #4} 
\def\PBphreva #1:#2 #3 #4 {#1,#2, {Phys. Rev. A} \PBvp #3 #4} 
\def\PBphs #1:#2 #3 #4 {#1,#2, {Physica Scripta} \PBvp #3 #4} 
\def\PBjqsrt #1:#2 #3 #4 {#1,#2, {J. Quant. Spectrosc. Radiat. 
       Transfer} \PBvp #3 #4} 
\def\PBcjp #1:#2 #3 #4 {#1,#2, {Can. J. Phys. } \PBvp #3 #4} 
\def\PBjphb #1:#2 #3 #4 {#1,#2, {J. Phys. B} \PBvp #3 #4} 
\def\PBapop #1:#2 #3 #4 {#1,#2, {Appl. Opt.} \PBvp #3 #4} 
\def\PBgca #1:#2 #3 #4 {#1,#2, {Geochim. Cosmochim. Acta}\PBvp #3 #4} 

\def\PBs{\phantom{n}} 
%
%
\def\PBkms{$\rm km s^{-1}$}

\def\PBt{${T}_{\rm eff}~$} 
\def\PBg{$\rm \log g$} 
\def\PBal{{ et al.~}}

%
%
%
%
%
\newcount\PBtn 
\def\PBcleartn{\global\PBtn=0} 
\def\PBtbl #1{\global\advance\PBtn by 1 
\begintable{\the\PBtn} 
\caption{{\bf Table \the\PBtn .} #1} 
} 
\def\PBtbltwo #1{\global\advance\PBtn by 1 
\begintable*{\the\PBtn} 
\caption{{\bf Table \the\PBtn .} #1} 
} 
\input psfig 
 
\begintopmatter  
 
\title{Abundances of  metal-weak thick-disc candidates} 
\author{ 
 P. Bonifacio,  M. Centurion and P. Molaro } 
\affiliation{Osservatorio Astronomico di Trieste, Via G.B. Tiepolo 11 
34131, Trieste -- Italy}

\shortauthor{P. Bonifacio, M. Centurion and P. Molaro } 
\shorttitle{Thick disc} 
 
 
\PBcleartn 
 
\abstract { 
High resolution  spectra of 5 candidate metal-weak thick-disc 
stars 
suggested by Beers \& Sommer-Larsen (1995) 
are analyzed to determine their chemical abundances. 
The low abundance of all the objects has been confirmed 
with metallicity reaching [Fe/H]$=-2.9$. 
However, for three objects, the astrometric data from the Hipparcos catalogue 
suggests they are true halo members. 
The remaining two, for which  proper-motion data are not available, 
may have disc-like kinematics.  
It is therefore clear that  it is useful to 
address  properties 
of putative metal-weak thick-disc stars
only if they possess full kinematic data. 
For CS 22894-19 
the abundance pattern similar to those of typical 
halo stars is found, suggesting that chemical composition is not 
a useful discriminant between thick-disc and halo stars. 
CS 29529-12 is found to be C enhanced 
with [C/Fe]=+1.0; other chemical peculiarities 
involve the $s$ process elements:  
[Sr/Fe]$=-0.65$ and [Ba/Fe]=+0.62,  leading to a high [Ba/Sr] 
considerably larger than what is found in 
more metal-rich carbon-rich stars, but similar to 
LP 706-7 and LP 625-44 
discussed  
by Norris \PBal (1997a). 
Hipparcos data have been used to calculate the space velocities of 
25 candidate metal-weak thick-disc stars, thus allowing us to  
identify 3 {\it bona fide} members, 
which  support the existence of a metal-poor tail of the thick-disc, 
at variance with a claim to the contrary by Ryan \& Lambert (1995). 
} 
 
\maketitle  

\section{Introduction} 
 
The disc of our Galaxy is 
rotationally supported  with a small velocity dispersion 
and a scale height of about 300 pc. 
The existence 
of  a ``thick disc'' was highlighted by Gilmore \& Reid (1983) 
using star count data. The proposed scale heights are 
in the range 1.0-1.5 kpc,  the asymmetric drift is in 
the range of 30-50 \PBkms, and the velocity dispersion is about twice 
that of the thin disc. 
Discs with a vertical 
scale height in excess of 1 kpc 
may be seen in  external edge-on spiral galaxies with small bulges, 
similar to our own 
(Van der Kruit \& Searle 1981a,1981b,1982). 
Because suggested kinematical and chemical parameters of the thick disc 
are similar  to those of the thin disc, this makes their determination 
prone to selection effects, and the issue of discreteness 
difficult to establish or reject. 
Whether the thick disc is a discrete component of 
the Galaxy or it is merely an ``extended'' configuration 
of the  disc, as suggested by Norris \& Ryan (1991), 
is still an open question.  For the sake of clarity, 
in the following we shall 
refer to the two as ``thin''(scale height $\approx$ 300pc) 
and ``thick'' (scale height $>$1 kpc) without implying by this 
the discreteness of the components. 
The origin of the thick disc is not clear. 
In their review Gilmore, Wyse \& Kuijken (1989) 
list the  possible formation mechanisms, 
which fall into three broad categories: 1) slow formation of the disc ; 
2) rapid formation of the thin disc and successive heating ; 3) accretion 
of satellite galaxies. 
 
The  thick disc is probably as old as the halo: 
Wyse \& Gilmore (1995) claim that the thick disc is ``apparently 
exclusively old''.  
The absence  
of RR Lyrae variables with disc kinematics 
would rule out that the disc 
is as 
old as the globular clusters, as had been noted by Majewski (1992). 
However 
the recent work of Martin \& Morrison (1998) 
using new kinematic data for a sample of 130 nearby RR Lyrae variables, 
was able to isolate a subsample of 21 stars with [Fe/H] $\le -1.0 $ 
and disc-like kinematics. 
The work of Fuhrmann (1998a), which makes use of Hipparcos data and 
high resolution, high S/N spectra, of about 50 nearby F and G type stars,
supports the notion  
that the thick disc is as old as the halo 
($\approx 12 $Gyr), while 
the thin disc is younger than 9-10 Gyr, thus highlighting 
a hiatus in star-formation in the Galaxy. 
 
\par 
The existence of the thick disc  and its kinematical parameters 
have a direct 
bearing on the interpretation of data aimed at the derivation of 
the rotational properties of the halo. As was pointed out by 
Majewski (1992) `` halo tracer surveys select stars from regions 
dominated by either the thin or thick disk''. 
A mixture of rapidly rotating disc stars and retrograde rotating 
halo stars could explain why such surveys determine a low prograde 
rotation for the halo, while Majewski's high $z$ sample reveals a 
sizeable retrograde motion. 
 
The kinematical distinction 
between disc (thin or thick) and halo is rather abrupt 
and should not give rise to any confusion. 
However, if we consider a metallicity biased sample of faint stars, 
their kinematics are imperfectly known (in general only radial velocity 
and photometric or spectroscopic parallax are available), making it  
impossible to assign unambiguous membership to the disc or to the halo 
on kinematical grounds. Can a selection may be made 
on chemical grounds alone? An answer to this question 
requires knowledge of the metallicity distribution of the 
thick disc. The metallicity of the thick disc is 
considered to be ``intermediate'' between 
thin disc and halo 
(Gilmore \& Wyse, 1985; Carney Latham \& Laird 1989) but 
a well defined  metallicity distribution is not 
yet available. 
Given the difficulty of isolating a pure thick disc sample,
it is not surprising that different 
selection criteria lead to different answers. 
Several studies have found a mean metallicity 
of the thick disc between $-0.6$  and $-0.4$  
(Gilmore \& Wyse 1985, Sandage \& Fouts 1987; Yoss, Neese \& Hartkopf 1987, 
Carney \PBal 1989; Friel 1987, Hartkopf \& Yoss 1982; Ratnatunga \& Freeman 
1989). 
Morrison, Flynn \& Freeman (1990) claimed that a metal-weak tail of the 
thick disc extends down to [Fe/H]$=-1.6$. 
On the other hand, Majewski (1992) finds a mean metallicity of 
-0.13, but suggests that the distribution is markedly 
non-gaussian   and has a metal weak tail even more extreme than suggested 
by Morrison \PBal (1990). 
Gilmore \& Wyse (1995) find from their 
{\it in situ} sample that the metallicity distribution peaks at 
[Fe/H]$=-0.7$ and there is a large overlap in metallicity 
between thin and thick disc, the thin disc extending down to 
at least -0.75 and the thick disc up to at least $-0.3$. 
Reid \PBal (1997), studied a sample of faint M dwarfs at high galactic 
latitude and found their data to be consistent with star-count models 
including three components (thin disc, thick disc, halo) and that up to 
2 kpc above the plane the abundances are close to the values of the 
old thin disc. 
The analysis of Reid (1998), who used a recalibration of 
the colour magnitude diagrams of globular clusters based 
on local subdwarf calibrators with Hipparcos 
data, to define a reference grid in the ($M_V,B-V$) plane, 
implies that $\approx$ 15\% 
of local disc stars have [Fe/H] below $-0.3$. 
Beers \& Sommer-Larsen (1995) suggest that stars with disc-like 
kinematics may be found for arbitrarily low abundances, although their 
sample, selected mainly on the criterion of low metal abundance, 
is not well suited to determine the metallicity distribution of 
the thick disc. 
In their paper Beers \& Sommer-Larsen provided a list of likely  
metal-poor thick disc 
stars, i.e. stars whose position is such that the line of sight 
component of the thick disc rotational velocity is in excess of 100 
\PBkms.  

Given the probable similarity in age of the halo and thick-disc, the 
existence of a metal-weak tail should be expected. 
However the existence of a metal-weak tail of the thick disc 
is seriously challenged by the results of 
Ryan \& Lambert (1995), who analyzed at high spectral resolution 
a sample of expected metal-poor disc stars taken from the 
samples of Norris \PBal (1985) and Morrison \PBal (1990) and found 
no star with disc kinematics below [Fe/H]$=-1.0$, arguing  
that the metallicities inferred from DDO 
photometry by Morrison \PBal (1990) 
were underestimated.

Nissen \& Schuster (1997) studied  a sample of stars 
of intermediate metallicity belonging to both the disc 
and halo  for which their lowest metallicity disc star 
has [Fe/H]=$-1.3$. 
In the 
Fuhrmann (1998a)  sample  
the most metal weak 
thick disc 
star is around [Fe/H]$=-1.0$.  
He was able to show 
that a discrimination between thin and thick disc is possible, 
if a combination of kinematics, abundance and age data is used. 

 Here we report on the detailed abundance 
analyses of 5 of the most interesting 
candidates, based on high resolution spectra. Li abundances 
for two of these objects have already been reported 
by Molaro, Bonifacio \& Pasquini (1997).

\section{Observations and data reduction} 
 
In September 1996 we observed 
CS 29529-012,CS 22894-019,BD $-14^\circ 5890$, HD165195,
HD26169 and  CS 22959-007
with the CASPEC 
spectrograph at the ESO 3.6 m telescope.  
The width of the lines of CS 22959-007 
revealed that it is a spectroscopic binary, it has been therefore 
dropped from the present analysis and we plan to collect other 
observations to elucidate its nature. 
The stars which were selected
from Table 8 of Beers \& Sommer-Larsen, 
had already been observed by us in the red (Molaro Bonifacio \&
Pasquini 1997).
The echelle grating with 31.6 
lines mm$^{-1}$ and the Long Camera (f/3) were used.   
The detector was the Tek CCD 
$1024\times 1024$   
with 24 micron wide pixels and a readout noise of about  
4e$^-$ per pixel. 
With the Long Camera the scale at the detector was about 0.059 {\AA} /pixel 
at 473.6 nm. 
The slit was set at $1.2^{''}$ arc seconds providing a resolving power of 
$ R = \lambda / \delta\lambda$ $\approx$ 30000, as measured from the 
Thorium-Argon emission lines.  The echellograms were reduced using the {\tt 
MIDAS ECHELLE} package.  The background was subtracted and cosmic rays removed 
with the {\tt FILTER/ECHELLE} command.  The images were not flat-fielded since 
this process increases the noise in the spectra and, at the wavelengths covered 
by our data, fringing is not important.  The spectra were then wavelength 
calibrated using the Th-Ar lamp spectrum, the root mean square error 
of the calibration was of the order of 0.2 -- 0.3 pm for all the 36 
orders observed on the detector. 
The spectrum of CS 29529-012 and one of the spectra of 
CS 22894-019 were taken in bright time and the sky spectrum was 
subtracted. 
The other spectrum of CS 22894-019 was taken during a total moon eclipse 
 and  sky subtraction was not necessary. 
The spectra 
of the other significantly brighter stars   
were not seriously affected by the sky,  
because the exposure times were much shorter . 
For CS 29529-012 and CS 22894-019 we also made use of the red 
spectra described in Molaro, Bonifacio \& Pasquini  (1997). 
 
A synthetic spectrum was computed for CS 29529-012  
and CS 22894-019 using the atmospheric 
parameters derived by Molaro \PBal (1997) and over-plotted  on 
the observed spectrum for identification purposes.  
Blended or noisy lines were removed 
in order to define a 
preliminary list of {\it bona fide}  lines to be measured. 
These lines were measured on all the available spectra by fitting 
a gaussian with  the 
IRAF task {\tt splot} which allowed the determination of line positions 
and equivalent widths.  Since
the dark-time spectrum of CS 22894-019 was by far superior 
to the bright-time one, we  used only the equivalent 
widths measured from this one. However the line positions were 
taken from both and the resulting radial velocities were averaged.

\section{Atmospheric parameters} 
 
For the  stars BD $-14^\circ 5890$, HD165195 and  HD26169 
we adopted effective temperatures from the literature, as detailed 
in Table 1, while for   CS 29529-012  and CS 22894-019 
 we adopted the temperature 
derived from $(B-V)_0$ by Molaro \PBal (1997). 
 
For all the stars the gravity was derived from 
the FeI/FeII ionization equilibrium and the microturbulent 
velocity  by the constraint that the FeI abundance 
be independent of equivalent width. 
Given the poor S/N of the spectra of the CS stars we were 
forced to use mainly strong lines and therefore the  microturbulent 
velocity is poorly 
constrained.

Both CS 22894-19 and CS 29529-12 are 
classified as dwarfs by Beers \PBal (1992). 
For CS22894-19 
we derived a gravity of log g = 3.5 using  3 Fe II lines. 
We had excluded the FeII 516.9 nm line because 
it yields an abundance which is almost 0.5 dex lower than that of the 
other three lines,  regardless of the gravity. 
By taking the 
Fe II abundance to be the average of 
all four lines one would derive a gravity of log g = 4.0. 
Since we assumed this line to be affected by noise we removed it 
from the analysis. 
Further support for the subgiant status of CS22894-19 comes from 
the Mg I b triplet lines: with log g = 3.5 the computed 
lines provide a satisfactory fit, whereas 
with a log g = 4.0 the computed lines are considerably broader 
than the observed ones.

Our gravity 
for CS29529-12 was log g = 3.0 derived using 
 five Fe II lines. Both Fe I and Fe II 
have a rather small rms scatter of about 0.2 dex in the abundances. 
Since a change  
as large as  0.5 dex in log g does not produce a difference larger 
than 0.2 dex  in the abundances of the two species, 
we  consider the gravity to be uncertain by $\pm 0.5$ dex. 
With the log g =3.00 the computed line strengths 
of the b triplet  are internally consistent 
and 
match  
the observed one.  
However the Mg I 470.2991 nm line yields an abundance which is 
about 0.5 dex larger.  A decrease of surface gravity 
or microturbulent velocity lessens the 
disagreement among Mg lines. For log g = 2.5 and $\xi=1.5$, values 
which are not unreasonable, the difference is 0.37 dex, 
and for log g = 2.0 $\xi=1.5$ it is 0.27 dex. 
This is because the Mg b triplet lines are sensitive 
both to pressure and to microturbulence, while 
the abundance from the Mg I 470.2991 nm line is not very sensitive 
to gravity, where  a decrease of 1 dex in log g produces only an increase 
of 0.065 dex in Mg abundance. 
 
Recently it has been realized that ionization equilibria provide 
gravities which are up to 0.5 dex lower than the 
gravities  derived in stars with accurate Hipparcos parallaxes 
(Nissen Hoeg \& Schuster 1997;  
Fuhrmann 1998b).  
We  preferred the log g given by ionization equilibrium 
for abundance analysis, both for homogeneity with the analysis 
of the two stars for which Hipparcos data are not available, and 
because  
the structure of the atmospheric  model  
satisfying the iron ionization equilibrium better 
reproduces 
the observed spectrum than does a model with the Hipparcos 
gravity.

\PBtbl{Basic data of program stars} 
\halign to \ColumnWidth{ 
\tabskip=15ptplus32ptminus15pt 
#&#&#&#&#&#&#&\hfill#\cr 
\multispan{8}{\hrulefill}\cr\cr 
Star & \PBt & \PBg & \PBg & [Fe/H] & [M/H]$^a$ & $\xi$ & $v_r$\cr 
     &      & spec.& Hip. & & & kms$^{-1}$ & kms$^{-1}$\cr 
CS 22894-19 & 6060 & 3.5 & - & -2.90 & -3.00 & 2.0 & $32 \pm\phantom{1} 2$\cr 
CS 29529-12 & 6197 & 3.0 & - & -2.27 & -2.00 & 1.7 & $93 \pm\phantom{1}  2$\cr 
BD $-14^\circ 5890$ & 4700$^b$ & 1.4 & 2.34  & -2.52 & -2.50 & 1.9& $ 118 \pm\phantom{1}  2$\cr 
HD 165195   & 4500$^c$ & 1.1 & 1.42  & -1.92 & -2.00 & 1.6& $0 \pm\phantom{1}  2 $\cr 
HD 26169    & 5000$^d$ & 2.0 & 2.46 & -2.61 & -2.50 & 1.9 & $20 \pm 15$\cr 
\cr 
\multispan{8}{\hrulefill}\cr\cr 
\multispan7{$^a$ metallicity of the adopted model\hfill}\cr 
\multispan7{$^b$ Cavallo et al (1997)\hfill}\cr 
\multispan7{$^c$ Francois (1996)\hfill}\cr 
\multispan7{$^d$ Peterson et al (1990)\hfill}\cr 
} 
\endtable

\subsection{Evolutionary status} 
 
To gain some insight into the evolutionary status of the 
stars in our sample we compared their temperatures and gravities 
with those predicted by theoretical isochrones. 
In Figure 1 we show  the \PBt - \PBg ~ diagram with  
the Revised Yale Isochrones (RYI ; Green et al 1987) for 
metallicity [Fe/H]$=-2.0$ and $-3.0$, an age of 12 Gyr and  
helium mass fraction Y=0.25. 
The isochrones of the 
Padova group (Bertelli et al 1994)  
are also shown for [Fe/H]=-1.7 (z=0.0004), 
ages of 12 Gyr and 6 Gyr, and Y=0.23. 
 
\beginfigure{1} 
\psfig{figure=fig2.epsi,width=8.4cm,clip=t} 
\caption{{\bf Figure 1.}  
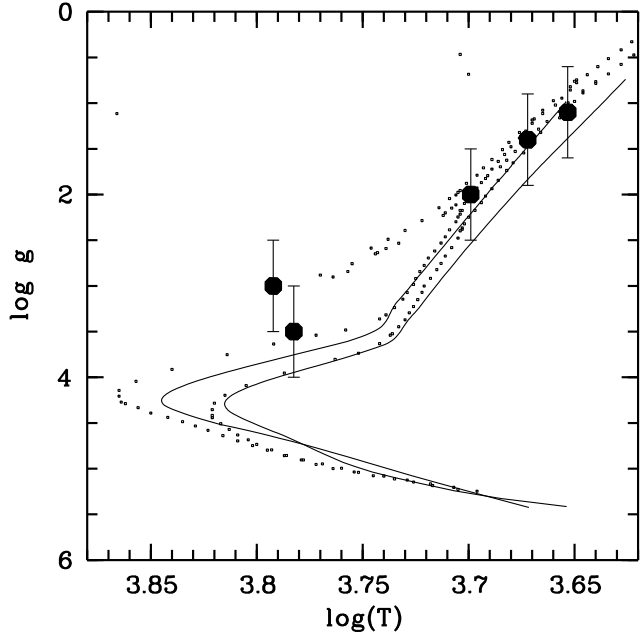
\PBt-\PBg diagram for our program stars. Solid lines 
are RYI isochrones for [Fe/H]=$-2.0$ and $-3.0$ (bluemost), 12 Gyr 
and Y=0.25. The dots represent the Padova isochrones 
for [Fe/H]$=-1.7$, 12 Gyr and 6 Gyr (bluemost) and Y=0.23.  
The age-metallicity degeneracy is evident. 
} 
\endfigure 
 
The position of the two CS stars is anomalous, although given 
the large uncertainty associated with their gravities they are in fact 
compatible with a TO or SG status. They actually occupy 
an area in the diagram occupied by Horizontal Branch stars, or 
Subgiant stars of much younger age (below 8 Gyr).  
For the three giant stars, for which Hipparcos parallaxes are 
available, we may compute gravities using the equation 
$$ \log {g\over g_\odot} = \log {M\over M_\odot} + 
0.4{(V_0+B.C.)}+2\log\pi +0.12$$ 
(see e.g. Nissen et al 1997), where we assumed $M=0.8M_\odot$, 
as is appropriate for giants stars on the RYI isochrone 
for  [Fe/H]=$-2.0$, shown in figure 1 (note that changing the mass 
to 0.5$M_\odot$ would decrease the \PBg ~ by only 0.2 dex). 
These gravities are, as expected, larger than the spectroscopic 
gravities and are given in table 1. 
The HB hypothesis 
may be discarded, because both  stars show a normal  Li  
abundance ( Molaro \PBal 1997) 
while 
a star during its red giant phase is expected to decrease 
its photospheric Li content due to depletion and dilution . 
That the stars are very young is rather problematic considering  
their low abundance. Moreover the results of Fuhrmann (1998a) 
imply that the thick disk is an old population.  
From our spectroscopic gravities we may derive distances 
for these stars, which are much greater than those 
given by Beers \PBal (1992), who assumed a TO status. 
For CS22894-19 we derive 1.7 kpc, compared to 1.1 kpc of 
Beers \PBal; for CS29529-12 1.2 kpc, compared to 0.4 kpc 
of Beers \PBal . 
 
It is for CS 29529-12 that 
the situation is more critical: 
if the ionization equilibrium  
gravity is wrong  
then  the star can really be  a turn-off star. 
If this is the case,  then 
the   surface gravity is around log g = 4.0, whichever isochrone 
set we use. However, on 
performing an abundance analysis with this surface gravity 
we find 
several chemical  inconsistencies, as shall  
be detailed in section 5.  
There will be a discrepancy of  0.34 dex between 
the results for Fe I and Fe II, which 
is not inconsistent with the rms scatter of each species.  
The chemical composition would then show some  
peculiar  ratios: [Mg/Fe]=$+0.17\pm 0.27$  [Ca/Fe]=$+0.13\pm 0.26$ 
and [TiII/FeII]=$+0.40\pm 0.29$.
The errors are estimated by just summing under quadrature
the line to line scatter given in Table 3.
The errors in \PBt and $\xi$ for [Mg/H], [Ca/H], [Ti/H] and [Fe/H]
are in the same direction and about the same magnitude, 
so that they will cancel out. The errors in \PBg,
instead, are in different directions for [Mg/H], [Ca/H] and [Fe/H],
so that this contribution should also be added under quadrature,
making the errors on [Mg/Fe], and [Ca/Fe] even larger.
For Ti we considered only the [TiII/FeII] so that errors in
\PBg ~ will cancel out.
Although 
the errors allow all these ratios to be consistent with a ``standard''
[$\alpha$/Fe]$=+0.4$, their values, taken at face value, appear
anomalous.
While the existence of $\alpha$-poor metal-poor stars has 
been reported by Carney et al (1997), Nissen\& Schuster (1997) 
and Jehin et al (1997), 
it is difficult to believe that Ti is over-enhanced 
with respect Mg and Ca. In fact while Mg and Ca are ``pure'' 
$alpha$-elements, Ti is also produced in nuclear statistical 
equilibrium together with iron-peak elements, thus it is expected 
to be under-enhanced with respect to Mg and Ca 
(Timmes \PBal 1995). 
Observers have either claimed that 
Ti goes in lockstep with Ca 
while Mg is over-enhanced (Primas \PBal 1994, Ryan \PBal 1996), 
or that Ti is slightly under-enhanced with respect to 
Ca and Mg (McWilliam \PBal 1995). So far there has been no report 
of a  Ti over-enhancement 
in metal-poor stars.

\section{Model atmospheres and analysis} 
 
For each star we computed a model-atmosphere using 
version 9 of the ATLAS code (Kurucz 1993) with the above 
determined parameters. We used the $\alpha$-enhanced opacity 
distribution functions (ODFs),  a microturbulence velocity of 1 \PBkms 
and no convective overshooting. These model atmospheres 
and the measured equivalent widths were used as 
input to the version 9 of the WIDTH code (Kurucz 1993) to 
determine the elemental abundances. At this stage some of the lines 
were removed from the analysis when their abundance 
was deviant from that of the other lines of the same species by more than 
1 $\sigma$. 
Equivalent widths, atomic data and derived abundances 
for the lines which were  retained are listed in 
Table 2.   
 
We consider errors  which arise from 
errors in the equivalent widths and errors in the model-atmosphere 
parameters. 
We ignore  errors in the atomic data and errors 
due to shortcomings in the 
model-atmosphere 
(treatment of convection, reliability of 1D-models). 
Formula (7) of 
Cayrel (1988) 
provides a conservative estimate of the errors in the equivalent widths: 
for our instrumental setup,  we estimate an error 
of  0.5 pm for the dark-time spectrum 
of CS22894-019,   0.8 pm for the spectrum 
of CS29529-012 
and around 0.1 pm for the giants. 
Misplacement of the continuum can  introduce an 
error in the equivalent width of up to  1.6 pm.
 Somewhat smaller errors are 
 expected for the lines which could be measured 
in two adjacent orders. 
We believe that a realistic estimate of abundance errors may 
be obtained from the line to line scatter when several 
lines are available. 
We investigate in detail the errors only for CS29529-012. 
For each element in Table 3 we report five errors:    $\sigma$ is 
 the r.m.s. 
error or line to line scatter for species with more than one 
line, the other columns are the variation in abundance deduced 
for an increase in log g by 0.5 dex, a decrease in \PBt by 130 K, 
an increase in $\xi$ by 0.3 \PBkms and an increase in equivalent 
widths by 0.4 pm. 
These errors may be used as a guide to assess the influence on 
the abundances of errors in model parameters and equivalent widths. 
We expect the situation for the other stars to be quite similar.

\section{Results} 
 
For  
CS 22894-19 and CS 29529-12 we derived 
[Fe/H]$=-2.90$ and $-2.27$ respectively,  
the metallicities derived from the Ca II K index 
by Beers \PBal (1992) for the two stars are 
$-3.03$ and $-1.54$, respectively. 
While for CS 22894-19 there is a substantial agreement, 
for CS 29529-12 our metallicity is considerably lower than  
previously estimated. 
A stronger than normal Ca II K line, or contamination
of an interstellar Ca II line, could have explained 
the discrepancy between our metallicity and that 
of Beers \PBal (1992). However 
we verified with spectrum synthesis that 
the Ca abundance derived from the Ca I lines is compatible 
with the Ca II H and K lines. 
 
For a few lines in the giant stars we resorted to spectrum synthesis 
either because the lines are mildly blended or because we felt 
that the gaussian fit used to measure the equivalent width 
did not capture satisfactorily the line wings  as is 
the case for the Mg I b triplet lines of HD 161695. 
 
For Mn we report the abundances only for the two CS stars, because
for the giants the lines are saturated,  and thus not suitable 
for abundance determination. The lines were 
treated by spectrum synthesis accounting for  the hyperfine 
splitting. The relevant atomic data 
was taken from McWilliam \PBal (1995). 
 
\PBtbltwo{line data and individual abundances} 
\halign to \PageWidth{ 
\tabskip=15ptplus32ptminus15pt 
#& 
\hfill $#$& 
\hfill $#$& 
\hfill $#$& 
\hfill $#$& 
\hfill $#$& 
\hfill $#$& 
\hfill $#$& 
\hfill $#$& 
\hfill $#$& 
\hfill $#$& 
\hfill $#$& 
\hfill $#$& 
\hfill $#$ 
\cr 
\multispan{14}{\hrulefill}\cr 
\hfill  \hfill & \lambda (\rm nm) & \chi (\rm cm^{-1}) & \rm log gf 
& EW (pm) & \epsilon 
& EW (pm) & \epsilon 
& EW (pm) & \epsilon 
& EW (pm) & \epsilon 
& EW (pm) & \epsilon \cr 
 & & & & 
\multispan2{\hfill {\rm CS 22894-19}\hfill}& 
\multispan2{\hfill {\rm CS 29529-12}\hfill}& 
\multispan2{\hfill \rm BD $-14^\circ 5890$\hfill}& 
\multispan2{\hfill \rm HD 165195\hfill}& 
\multispan2{\hfill \rm HD 26169\hfill} 
\cr 
\multispan{14}{\hrulefill}\cr 
Na I & 588.9951& 0.000  & 0.117 & 6.57 & 3.15  & 19.10 & 5.57  \cr 
Na I & 589.5924& 0.000  & -0.184& 4.47 & 3.13  & 17.20 & 5.68   \cr 
Mg I & 470.2991& 35051.264 & -0.666 & 2.61 & 5.33 & 7.36 & 6.21& 9.72 & 5.92 & -     & -6.20 & 8.08 &  5.76\cr 
Mg I & 517.2684& 21870.464 & -0.402 & 11.26 & 4.87& 14.50& 5.71& 27.00 & 5.51& -     & -6.40 & 21.12 & 5.31 \cr 
Mg I & 518.3604& 21911.178 & -0.180 & 13.20 & 4.92& 15.60& 5.63& 30.56 & 5.43& -     & -6.40 & 24.52 & 5.31\cr 
Al I & 396.1520& 112.061 & -0.323   &  6.98 & 3.01&  9.28& 3.70& 28.25 & 3.14& 22.82 & 3.90  & 13.62 & 3.16 \cr 
Ca I & 422.6728& 0.000   & 0.243    & 10.75 & 3.49& 12.63& 4.26& 30.02 & 4.15&  -    &  -    & 23.13 & 4.12\cr 
Ca I & 430.2528& 38551.558& 0.275   & 4.48  & 4.01&  5.69& 4.37& -    &  -   &  -    &  -    & -  &  -\cr 
Ca I & 445.4779& 37757.449& 0.252   & 3.84  & 3.91&  5.50& 4.35&  8.98 & 4.05&  -    &  -    & 7.96  & 4.01\cr 
Ca I & 616.2173& 31539.495& 0.100   &  -    & -   &  6.52& 4.65&  -    & -   &  -    &  -    &  -    &  -\cr 
Sc II & 424.6822& 2540.950& 0.322   & 3.49  & 0.07&  7.24& 0.74& 12.51 & 0.42&  15.83&  1.12 & 10.89 & 0.44\cr 
Ti II & 430.0049& 9518.060& -0.750  & 3.75  & 2.39&   -  &  -  &  -    &  -  &   -   &   -   &  -    &  - \cr 
Ti II & 439.5033& 8744.250& -0.650  & 5.00  & 2.41&  9.30& 3.31&   - &  -    &  21.34& 3.89  &  10.17& 2.47\cr 
Ti II & 444.3794& 8710.440& -0.810  &  -    & -   &  6.80& 2.87&  11.22& 2.53&  14.59& 3.33  &  10.18& 2.63\cr 
Ti II & 446.8507& 9118.260& -0.600  & 3.90  & 2.21&  7.10& 2.76&  12.65& 2.69&  16.90& 3.48  &  10.83& 2.61\cr 
Ti II & 450.1273& 8997.710& -0.750  &  -    &  -  & 8.30 & 3.17&  12.00& 2.69&  16.08& 3.50  &  10.07& 2.56\cr 
Ti II & 457.1968& 12676.970& -0.530 & 3.61  & 2.49& 7.00 & 3.07&  11.91& 2.93&  16.86& 3.82  &  10.17& 2.85 \cr 
Cr I  & 425.4332& 0.000    & -0.114 & 3.99  & 2.49&  -   &  -  &  11.66& 2.67&  16.10& 3.35  &  9.79 & 2.57 \cr 
Cr I  & 427.4796& 0.000    & -0.232 & 4.40  & 2.68&  6.55& 3.33&  12.15& 2.89&  17.07& 3.64  &  10.48& 2.86 \cr 
Cr I  & 428.9716& 0.000    & -0.365 & 2.18  & 2.36&  9.94& 4.35&   -   &  -  &   -   &  -    &   -   &  -    \cr 
Cr I  & 520.4506& 7593.160 & -0.170 & 1.81  & 2.90&  3.20& 3.41&   -   & 3.17&   -   & 3.50  &   -   &   -    \cr 
Cr I  & 520.6038& 7593.160 & 0.030  & 2.05  & 2.77&  5.50& 3.66&   -   & 2.78&   -   &  -    &   -   &  2.72     \cr 
Cr I  & 520.8419& 7593.160 & 0.160  & 2.13  & 2.66&  4.40& 3.31&   -   & 2.98&   -   & 3.50  &   -   &   -     \cr 
Mn I  & 403.0753^*& 0.000    & - & -  & 2.20&   -  &  -  &  - & 3.00  &   -   &  -    &  12.04&  2.87\cr 
Mn I  & 403.3062^*& 0.000    & - & -  & 2.15&   -  &  -  &  - &  - &   -   &  -    &  10.17& 2.50\cr 
Mn I  & 403.4483^*& 0.000    & - & -  & 2.20&      &     &    & 3.12 &       &       &       &    \cr 
\multispan{14}{\hrulefill}\cr 
\multispan{14}{$^*$ 
treated by spectrum synthesis with HFS 
splitting as given by McWilliam et al (1995)\hfill}\cr 
} 
\endtable

The abundances in CS 22894-19 are similar to those 
of typical halo stars of comparable metallicity 
and in that respect unremarkable. 
 
CS29529-12 deserves some additional remarks. 
Among the observed Cr lines 
there is some discrepancy. 
The abundance of Cr given in Table 4,  
has been derived from one resonance line 
of multiplet 1 ,427.5 nm  
(we excluded CrI 425.4 nm because it is rather 
large in our spectrum and we suspect it to be blended, possibly 
with a Cr II line, and 429.0 nm because it yields 
an abundance which is 1 dex higher than the other lines) 
and the three lines of multiplet 7. All the log gf 
values are computed from the 
transition probabilities of the compilation of Younger et al (1978). 
The data for multiplet 7 are less accurate, owing to a possible 
self-absorption in the emission-line experiments; however their 
uncertainties should be within 50\% , which translates into 
a 0.2 dex uncertainty in log gf and therefore abundance.

The Sr abundance is derived from 
the Sr II 421.5 nm resonance line only, because the  
stronger Sr II 407.8 nm line 
is affected by noise in our spectrum. 
The barium abundance is obtained from the 445.5 nm line.  
This line may be potentially affected 
by hyperfine structure ({\it hfs}) 
but in an amount which depends on the relative 
importance of {\it s} or {\it r} processes (see Norris et al 1997a). 
When the {\it s} process is    
dominant there is a  
minimal production of the 
odd isotopes where the {\it hfs} is important. 
 
Sr is over-deficient by $-0.65$ dex, while 
Ba is over-abundant by 0.62 dex. 
This yields [Ba/Sr]=+1.3, which is rather 
unusual for ``normal'' metal-poor stars, where 
it is found negative.      
However $s-$ enriched metal-poor stars, such as 
carbon-rich stars (CH,sgCH), do show [Ba/Sr] 
ratios in the range 0.5 to 2, this is generally 
understood to be due to the fact that at low 
metallicities the main neutron source necessary 
for the $s-$ process is $^{13}$ C$(\alpha, n)^{16}$O,  
rather than $^{22}$Ne$(\alpha,n)^{25}$ Mg 
(Luck \& Bond 1982). 
In fact CS  29529-12  is mildly C enhanced 
with [C/Fe] on the order of $+1.0$. 
The carbon abundance has been derived from spectral 
synthesis of the G band. 
The star has not been noticed to have extremely strong G-band by  
Beers \PBal (1992),  
as  
its GP index  is 0.92, 
which is only slightly larger  
than the median GP value for metal-poor stars around metallicity -2.0. 
 
We do not have information on nitrogen and oxygen since the CN 388.3 nm band 
falls outside our  
spectral range and the noise at the OI 630.0 nm line 
does not allow us to place useful limits on the oxygen abundance. 
Adopting [O/Fe]=+0.6 dex which is typical of Population II stars 
we have  [C/O]=+0.4, which is not enough to classify it as a carbon star, 
but which clearly shows a carbon enhancement with respect to normal dwarfs. 
 
The cases of carbon-enhanced metal-poor  
stars are discussed  by Norris \PBal  (1997a)  
who emphasized the relatively high frequency of carbon-enhanced 
stars at low metallicity compared to current chemical 
galactic  evolution models. These stars show a large variety of 
chemical behaviour for  
the neutron capture elements going from the case of  CS 22892-52,  
where a large abundance of {\it r} elements is present, to the case 
of CS 22957-027 (Norris et al 1997b, Bonifacio et al 1998) where  
the C overabundance 
is observed together with a strong  
underabundance in the neutron-capture elements.

The distinctive peculiarity of CS 29529-12 is that the carbon enhancement 
is observed together with  a Ba  
enhancement, but with  very little Sr. 
The quantity [hs/ls],  
defined as the difference between the mean abundance  of 
the heavy $s$ process elements,  such as Ba, and the abundance of the 
lighter ones such Sr, is heavily weighted 
toward higher atomic number in stars which show neutron capture 
elements enhancement. 
 
Norris \PBal  found that the [hs/ls] ratio  increases at 
low metallicities being   
about 1.5 for the two stars LP 625-44 and LP 706-7  which have 
[Fe/H] around $-3.0$ . CS 29529-12 
is slightly more metal-rich and with a less extreme 
carbon abundance but an [hs/ls] ratio  very similar to that of these two stars. 
 
Chemical peculiarities in cool stars are interpreted to be 
the result either of 
mixing  in the star itself or of accretion from an evolved companion. 
In particular the formation of Ba stars, CH stars and dwarf carbon stars 
probably involves the evolution of a binary system. 
 
We further note, with regard to chemical peculiarities of CS 29529-12, that the 
lithium abundance  
($\epsilon(\rm Li)= \log (Li/H)+12 =2.21$) is found at the value typical 
of the other low-metallicity 
turn-off stars  
(Molaro et al 1997). If  
the star had undergone some unusual surface  
disturbance as a result of self-mixing or of the dredge up  
of processed material onto 
their surfaces,   
the Li surface abundance  would have been  
modified.  
Li is easily destroyed if mixed into deeper layers where temperatures  
reach a few million degrees. 
The apparent normality of Li is  
thus suggestive evidence that the star has not experienced major modifications 
of its surface abundance by mixing with internal layers. 
A similar case is that of LP 706-7 (= G268-32) with  
$\epsilon\rm(Li)=2.25$ (Thorburn 1994) 
and CS 22898-027 with $\epsilon\rm(Li)=2.52$ (Thorburn \& Beers 1992)

The binarity hypothesis remains favoured and should be checked by  
additional observations. 
While, for LP 625-44,   
radial velocity and Li abundance are  
consistent with the binarity  
for LP 707-7 there is no clear evidence of radial velocity 
variations and its Li abundance is at the  
the level of the {\it Spite Plateau}. 
It would be 
interesting to establish if CS 29529-12 is  analogous to LP 706-7,  
because these 
objects are very difficult to account for 
in present stellar evolution theory.

\beginfigure{2} 
\psfig{figure=fig_nai_disk.epsi,width=8.8cm,clip=t} 
\caption{{\bf Figure 2.}  
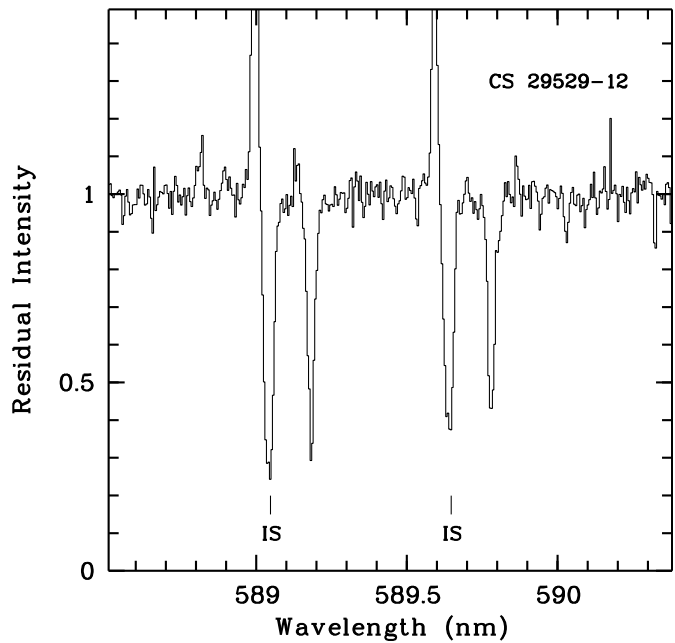
Na I D lines of CS 29529-12. 
} 
\endfigure

The abundance of Na deduced from the D lines is remarkably high, 
the lines are shown in Fig. 2. 
It is true that in this temperature and metallicity range 
the sodium D lines are known to be strongly affected by 
NLTE effects (Baum\"uller, Butler \& Geheren, 1998), however 
the correction should be at most of the order of -0.5 dex in [Na/Fe], 
thus  leaving an uncomfortably large overabundance of 1.0 dex.  
For confirmation of the overabundance we searched for the subordinate lines 
at 568.2633 nm, 568.82 nm and 616.0747 nm. The last  
is affected by atmospheric emission   
in our spectrum, the first two  
should have an equivalent width 
around 4 pm for the Na abundance implied by the resonance lines, 
one is marginally detected but the other is not, so the test 
is inconclusive. The sodium overabundance persists  
even if we adopt the gravity of log g = 4.0, although reduced 
by 0.6 dex. 
The three bright giant stars have been studied by several 
authors,  and so in the appendix 
we compare our results with the literature.

\section{Kinematics of metal weak thick disc candidates} 
 
For all the program stars we determined radial velocities from the 
lines used in the abundance analysis. The r.m.s scatter is 
on the order of 0.7-0.9 \PBkms ~for the brighter giant stars, 
while it is of the order of 1.6 \PBkms ~for the lower S/N data 
of the fainter CS  
stars. In both cases the resulting error in the radial velocity is dominated 
by the uncertainty in the zero point shift, which is on the order 
of 2 \PBkms, except for HD26169, for which an arc frame 
at that telescope position was not available and the zero point error 
could be as large as 15 \PBkms.  The radial 
velocities which we have determined from our spectra are in good 
agreement with those given by  Beers \& Sommer-Larsen 
(1995) and Molaro \PBal (1997), with the possible exception of HD26169, 
for which we note that while Beers \& Sommer-Larsen (1995) give $-24$ \PBkms , 
Norris (1986) gives $-35$ \PBkms. 
 
The three giants have parallaxes and proper motions measured with 
the Hipparcos astrometric satellite, but 
for the two CS stars the full kinematics cannot be recovered 
because there are no proper-motion data. 
 
We used the matrices given by Johnson \& Soderblom (1987) 
to compute spatial velocities and associated errors, adopting 
the left-handed reference frame in which U is directed 
towards the Galactic anti-center, V in the direction of Galactic 
rotation and W towards the North Galactic pole. We therefore 
changed the sign in the top row of the matrix given by 
Johnson \& Soderblom (1987). All the velocities given 
are heliocentric. 
The error in the space velocities is dominated by the error in the parallax. 
The proper motion 
and radial velocity data  yield errors in the 
derived space velocities of the order of a few \PBkms 
while the parallax data, yield errors which are over an 
order of magnitude greater. 
Therefore, we considered also the photometric parallax which we derived 
by estimating the absolute magnitude by interpolating in the 
proper RYI isochrone (Y=0.25, age of 12 Gyr and metallicity of the star) 
for the star's assumed effective temperature. 
These absolute magnitudes and the derived space velocities 
can be found in the Table 5. 
 
For BD $-14^\circ 5890$ the trigonometric parallax is formally 
undetermined because of its large 
error of 126\%. Its  space velocities derived 
from the parallax listed in the Hipparcos catalogue 
are quite ``normal'' halo velocities. On the other hand, the space velocities 
derived from the photometric  
parallax are those of   a runaway  
star,  with a large retrograde orbit, which is too extreme to be 
believable. 
Even if the parallax were  as large as 7 mas, i.e. 
3$\sigma$ larger than the Hipparcos parallax, 
the kinematics of the star would  still not be 
disc-like. Although its circular velocity would be similar to that of the sun 
its large $U$ and $W$ velocities would identify it as a halo member. 
We therefore consider  the Hipparcos parallax to be quite close to 
the true parallax in spite of the large error. 
For HD 26169 
we  exclude a disc-like kinematics, for the same reasons 
as for BD $-14^\circ 5890$. 
For HD 165195 the trigonometric and photometric parallax 
agree within 0.03 mas. The derived space velocities differ little and 
qualify this star as a member of the halo. 
Of the three giants, however, this is the one with the smallest $U$ and $W$ 
velocities, which may suggest a disc motion. 
In fact if we assume  the parallax to be larger 
by 2$\sigma$ we find its rotational velocity to be 
150 \PBkms and its kinematics 
may be effectively described as disc-like.  
HD 165195 could possess  disc-like kinematics if its parallax were some 
$2\sigma$ larger than the Hipparcos one, however the good coincidence 
between the photometric and trigonometric parallaxes make this event 
unlikely. 
 
We next investigated how many of the 44 stars given 
in Table 8 of Beers \& Sommer-Larsen (1995) have Hipparcos data. 
We found 23 objects.   
We removed two from our sample because they have 
a negative  parallax.   
 
The kinematics of HD 111721 is quite remarkable, being in retrograde 
motion with a speed in excess of 600 \PBkms . 
For the three RR Lyrae variables in this list the full space 
kinematics have already been computed by Martin \& Morrison (1998), using 
proper motions from Hipparcos combined with ground-based data, 
and distances from Layden (1994) which 
relied on an absolute magnitude calibration of the RR Lyrae variables. 
We used their data to compute the space velocities of these stars 
and confirm their conclusion that XX And and SW Aqr belong to the halo 
while EZ Lyr belongs to the metal weak thick disc. These data are also 
given in table 5. 
We draw essentially the same conclusion  if we use the radial 
velocities from Beers \& Sommer-Larsen (1995) 
and the Hipparcos parallaxes, except for EZ Lyr, whose 
Hipparcos parallax is undetermined (error larger than parallax). 
A caveat for RR Lyrae stars is that 
the radial velocity of a pulsating star is prone to systematic errors. 
We note here that the Simbad data base gives the values of -5 \PBkms 
and +21 \PBkms for SW Aqr and the values -25 \PBkms and -0.6 \PBkms for XX And . 
 
We also reconsider the kinematics of the stars of Carney \PBal (1997) since 
all of their ``candidate thick disc'' stars now have Hipparcos 
data. The space velocities and kinematical data 
are reported in Table 5. Two stars  (G023-014 and G190-015) show  
disc-like kinematics, two (BD +$80^\circ 245$ and G182-031) 
have halo kinematics while one, 
G090-003, shows  undetermined kinematics, since its Hipparcos 
parallax is undetermined (error larger than the parallax). 
 
On the basis of the spectroscopically determined gravity 
Carney \PBal computed space motions which are in substantial 
agreement with those given here for BD+80 245, G182-031 and 
G190-015, but in strong disagreement for G023-014.

By combining the Beers \& Sommer-Larsen and the Carney \PBal 
samples 
we compile  a sample of 
candidate metal-weak thick-disc stars which possess Hipparcos 
proper motions using the following criteria: 
we exclude  all stars whose Hipparcos parallax has an error 
greater than 200 \%, but include the RR Lyrae variables 
using the Layden (1994) distances and the Martin \& Morrison 
(1998) proper motions, which are based also on Hipparcos data. 
For these stars we adopted as errors on the space velocities 
(66,66,52) \PBkms, which is $\sqrt{22}\times\Delta$, 
where 22 is the size of their metal weak disc sample 
(sample DISK2A) and $\Delta$ are the errors of the mean 
velocities of the sample quoted in their table 3. 
This constitutes a sample of 25 stars for which absolute 
space motions can be computed. 
 
Figure 3 is a $U,V$ diagram for 
all those stars  and includes the three giants investigated in this paper. 
It can be seen that 
the majority of the stars qualify as true halo members. 
To select a  sub-sample of likely metal-weak thick-disc 
stars we must assume a specific 
set of kinematic parameters: we adopt the parameters 
of the sample DISK2A of Martin \& Morrison (1998) 
i.e. $(<U>,<V>,<W>)= (12 -59,-19)$ \PBkms 
and $(\sigma (U),\sigma(V),\sigma (W))=(64,64,62)$. 
As parameters we use $V$ and 
$K^2={ (U^2+W^2)}$, which is a measure of the kinetic energy 
not associated with  rotation. 
Clearly a star with high $K^2$ does not qualify as a disc star. 
To a good approximation we may write 
$<K^2> = \sigma^2(U)+<U>^2+\sigma^2(W)+<W>^2$ 
and $\sigma^2({K^2})={4<U>^2\sigma^2(U)+4<W>^2\sigma^2(W)}$. 
By imposing a $2\sigma$ criterion 
(we select all stars with 
 $K^2\in [2299,12311]\rm km^2s^{-2}$ 
and $V\in [-187,-69]\rm kms^{-1}$) 
we isolate 7 stars: G190-15, HD165195 
HD6446,      HD23592,   HD27928,   HD97320,    EZ Lyr. 
Only 3 stars match a $1\sigma$ criterion: 
G190-15, HD97320,    EZ Lyr. 
 
\beginfigure{3} 
\psfig{figure=fig1.epsi,width=8.8cm,clip=t} 
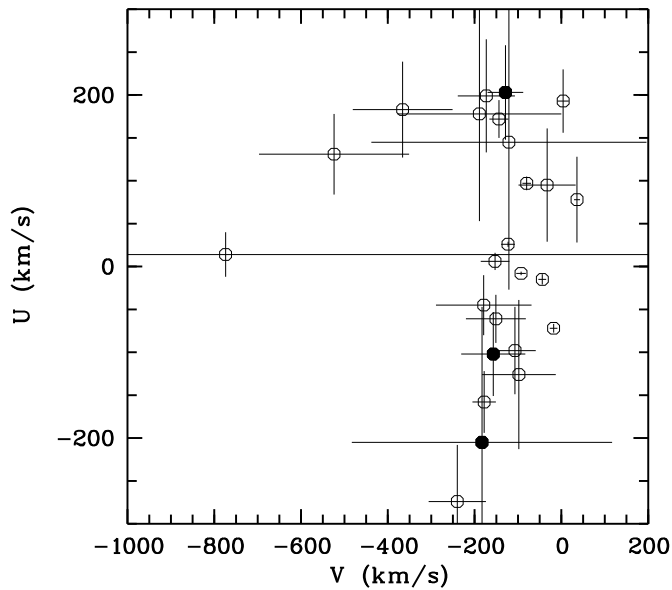
\caption{{\bf Figure 3.} $U,V$ diagram for our 
sample of stars, the three giants for which abundance 
analysis is reported in the paper are identified by  
crossed symbols} 
\endfigure 
 
With such a small sample very little can be said 
about the nature of the metal-weak 
thick-disc.  
One of the main conclusions is that Table 8 of Beers \& Sommer-Larsen 
(1995) is probably 
not the best place to look for metal-weak thick disc-stars. 
Quite likely the problem of finding such stars cannot 
be addressed with only incomplete kinematic information. 
In spite of this it has been possible to 
show that stars with disc kinematics and 
metallicity below [Fe/H]$\approx -1.0$ do exist. 
HD 166913 ([Fe/H]$=-1.80$) and HD 205650 ([Fe/H]=$-1.30$)  
are an other two metal-poor stars with disc kinematics 
which have been noted by Romano \PBal (1999) 
in the course of their analysis of the sample 
of stars with Li measurements. 
 
To demonstrate the situation we show a plot 
of [Fe/H] versus $K^2$ in Fig. 4. With such a diagram, 
once the effects of halo contamination 
have been understood, and with 
a significant number of stars one should 
be able to infer knowledge of 
the coupling of chemical and 
kinematical evolution of the thick disc. 
A uniform distribution of stars with [Fe/H] 
would support a slow formation of the 
thick disc, while a distribution 
peaked at some narrow range in [Fe/H] 
would testify in favour of a short 
timescale. 
 
\beginfigure{4} 
\psfig{figure=fig3.epsi,width=8.8cm,clip=t} 
\caption{{\bf Figure 4.}  
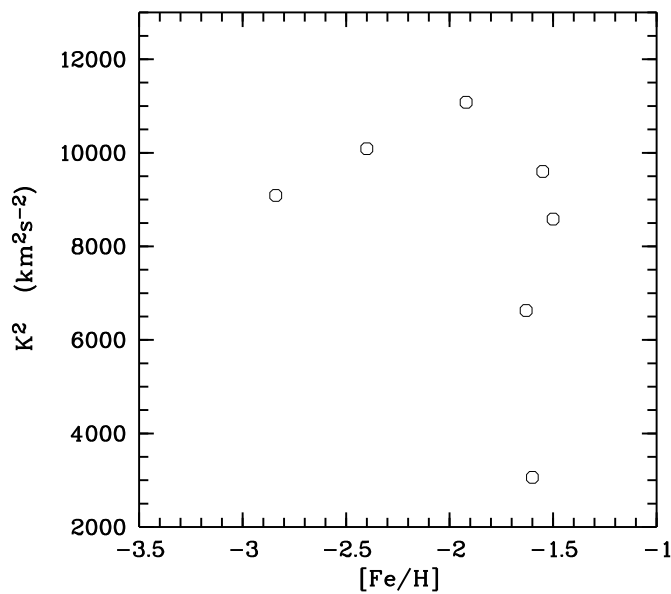
$K^2 (= U^2+W^2)$ as a function of [Fe/H]. 
Such a diagram  should help to infer knowledge 
on the coupling of chemical and dynamical 
evolution of the thick disc. 
} 
\endfigure 
 
\beginfigure{5} 
\psfig{figure=pl_bacs-12.epsi,width=8.8cm,clip=t} 
\caption{{\bf Figure 5.}  
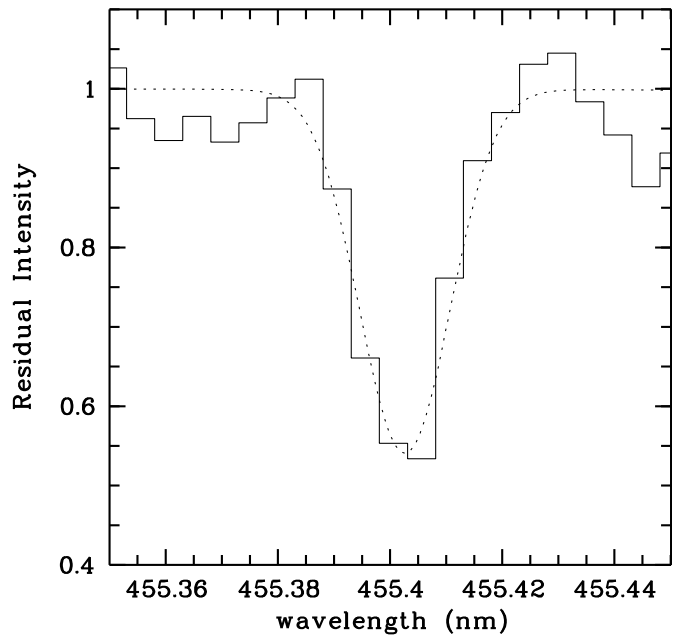
The Ba resonance line in CS  29529-12, the dotted line is 
a synthetic spectrum computed assuming  $\epsilon = +0.48$. 
} 
\endfigure

\section{Conclusions} 
 
Our analysis of 
chemical abundances has confirmed that all of  
our program stars are indeed metal-poor with  
iron abundances less than about one hundredth 
the solar abundance. 
The Hipparcos data  
allowed us  to compute the kinematics of 
the giant stars which qualifies them as halo members;  for 
the two fainter stars CS 22894-019 and CS 29529-012 
no proper motion data are available and  
their membership of the thick disk   
cannot be established or rejected. 
 
The ionization-equilibrium gravities  
are not compatible with a TO status of CS 22894-019 and 
CS 29529-012. On the other hand, their Li abundances 
prohibit their    interpretation  as HB stars. Therefore they 
are probably in the SG phase. 
 
The chemical composition of four out of  
our five program stars is that typical of halo 
stars. 
CS 29529-012 is C enhanced 
and has a high [Ba/Sr] ratio, 
which  features make it akin to the metal-poor carbon-rich 
stars studied by Norris \PBal (1997a). 
It also displays a huge Na abundance 
which, however,  
we regard as doubtful, 
in view of the non-detection of the subordinate Na I lines. 
Further observations are encouraged; it is  possible 
that the combined effects of NLTE and a higher gravity 
may result in a value of [Na/Fe] lower by about one dex. 
 
In order to shed further light 
on the metal-weak thick-disc, proper motion data 
for these two stars, as well as for the other 
metal-poor stars of comparable magnitude
are badly needed. 
The present paper, in line with other recent chemical analyses 
of thick disc stars (Nissen \& Schuster 1997, Fuhrmann 1998a), 
suggests that no chemical signature is likely to be found 
to discriminate between halo and metal-weak thick-disc.

\par 
 
\let\PBtn=\PBtn 
\begintable*{\the\PBtn} 
\caption{{\bf Table \the\PBtn .}(continued) 
{line data and individual abundances (continued)}} 
\halign to \PageWidth{ 
\tabskip=15ptplus32ptminus15pt 
#& 
\hfill $#$& 
\hfill $#$& 
\hfill $#$& 
\hfill $#$& 
\hfill $#$& 
\hfill $#$& 
\hfill $#$& 
\hfill $#$& 
\hfill $#$& 
\hfill $#$& 
\hfill $#$& 
\hfill $#$& 
\hfill $#$ 
\cr 
\multispan{14}{\hrulefill}\cr 
\hfill  \hfill & \lambda (\rm nm) & \chi (\rm cm^{-1}) & \rm log gf 
& EW (pm) & \epsilon 
& EW (pm) & \epsilon 
& EW (pm) & \epsilon 
& EW (pm) & \epsilon 
& EW (pm) & \epsilon \cr 
 & & & & 
\multispan2{\hfill {\rm CS 22894-19}\hfill}& 
\multispan2{\hfill {\rm CS 29529-12}\hfill}& 
\multispan2{\hfill \rm BD $-14^\circ 5890$\hfill}& 
\multispan2{\hfill \rm HD 165195\hfill}& 
\multispan2{\hfill \rm HD 26169\hfill} 
\cr 
\multispan{14}{\hrulefill}\cr 
Fe I  & 400.5241& 12560.930& -0.570 & 7.10  & 4.72&   -  &  -  &   -   &   - &   -   &  -    & 10.88 & 4.69\cr 
Fe I  & 404.5813& 11976.234& 0.220  & 10.17 & 4.58& 11.45& 5.35&   -   &   - &  -   &  -     & 17.46 & 4.92\cr 
Fe I  & 406.3594& 12560.930& 0.080  & 8.48  & 4.39& 8.92 & 4.86& 18.18 &4.92 & -   &  -      & 14.76 & 4.81\cr 
Fe I  & 407.1737& 12968.549& 0.000  & 8.25  & 4.45&  -   &  -  & 18.69 & 5.09&  -   &  -     & 14.44 & 4.90\cr 
Fe I  & 413.2058& 12968.549& -0.634 &  -    &  -  &  -   &  -  & 13.42 & 5.01 & 19.02 & 5.59 & 11.24 & 4.85\cr 
Fe I  & 414.3869& 12560.930&  -0.440& 6.58  & 4.47& 7.71 & 5.01&   -   &   - &  -   &  -     &   -   &  - \cr 
Fe I  & 418.7795& 43434.629& -0.520 & -     &  -  & -    &  -  & 10.71 & 5.22 & 12.85 & 5.69 & 8.53  & 4.95\cr 
Fe I  & 419.8304& 43163.327& -0.521 & 2.32  & 4.52& 5.52 & 5.35&   -   &   - &  -   &  -     & 9.25  & 5.09\cr 
Fe I  & 419.9095& 24574.650& 0.248  & 3.51  & 4.61&   -  &  -  & 9.96 & 5.02&  -   &  -      & 7.60  & 4.65\cr 
Fe I  & 420.2028& 11976.234& -0.697 & 5.94  & 4.53&  -   &  -  & 12.77 & 4.76 & 18.40 & 5.47 &   -   &  - \cr 
Fe I  & 423.3602& 43633.534& -0.560 & 2.67  & 4.72& 3.68 & 5.08& 9.73  & 5.06 & 13.39 & 5.86 & 8.50 & 5.02\cr 
Fe I  & 423.5936& 43163.327& -0.292 &  -    &  -  & 7.52 & 5.69&   -   &   - &  -   &  -     &   -   &  - \cr 
Fe I  & 425.0118& 43434.629& -0.370 & 2.69  & 4.52& 5.40 & 5.23& 9.23 & 4.73 & 12.04 & 5.38  & 7.90 & 4.67\cr 
Fe I  & 425.0787& 12560.930& -0.691 &  -    &  -  & 6.70 & 4.99& 12.55& 4.78 & 17.50 & 5.45  & 11.10 & 4.78\cr 
Fe I  & 426.0473& 42815.858& 0.129  & 6.22  & 4.61& 8.26 & 5.36&   -   &   - &  -   &  -     &   -   &  - \cr 
Fe I  & 427.1153& 19757.033& -0.334 & 3.63  & 4.65& 6.89 & 5.50& 10.26& 4.92 & 11.41 & 5.17  & 9.11 & 4.91\cr 
Fe I  & 427.1759& 11976.234& -0.127 & 8.19  & 4.42& 8.87 & 4.93&   -   &   - & 21.38 & 5.07&   -   &  - \cr 
Fe I  & 428.2402& 17550.175& -0.730 & 2.77  & 4.62& 5.45 & 5.33& 10.93 & 5.20&  -   &  -    & 9.25 & 5.04\cr 
Fe I  & 430.7902& 12560.930& -0.060 & 9.80  & 4.79&  -   &  -  &   -   &   - &  -   &  -    & - & - \cr 
Fe I  & 432.5762& 12968.549& 0.000  & 8.54  & 4.47& 10.73& 5.43 & 19.66 & 5.13&  -   &  -   & 14.60 & 4.85 \cr 
Fe I  & 437.5930& 0.000    & -3.031 & 1.74  & 4.58&  -   &  -   & 10.81 & 4.75& 16.23 & 5.86&  9.55 & 4.91\cr 
Fe I  & 438.3544& 12560.930& 0.160  & 10.37 & 4.62& 11.72& 5.48 & 19.68 & 4.79&  -   &  -   &  16.18& 4.76 \cr 
Fe I  & 440.4750& 12560.930& -0.180 & 9.49  & 4.82&  8.43& 4.91 & 17.68 & 5.03&  -   &  -   &  14.33& 4.89 \cr 
Fe I  & 441.5122& 12968.549& -0.580 & 7.40  & 4.79&  7.44& 5.09&   -   &   - &  -   &  -    &  11.62& 4.79\cr 
Fe I  & 452.8613& 17550.175& -0.822 & 3.87  & 4.91&  3.92& 5.10&   -   &   - &  -   &  -    & 9.83 & 5.16\cr 
Fe I  & 489.1492& 43434.629& -0.130 & 3.22  & 4.71&  5.12& 5.24& 10.59 & 5.09& 13.35 & 5.61 & 8.84 & 4.97\cr 
Fe I  & 492.0502& 43163.327& 0.070  & 3.30  & 4.51&  6.94& 5.41& 11.25 & 5.01& 16.80 & 5.89 & 9.87 & 4.97\cr 
Fe I  & 495.7298& 43163.327& -0.342 &  -    &  -  &  4.97& 5.42& 9.71  & 5.09& 18.14 & 5.90 & 7.54 & 4.90\cr 
Fe I  & 495.7597& 42815.858& 0.127 &   -    &  -  &   -  &  -  &  -    &  -  &  -    &  -   & 10.77& 5.09\cr 
Fe I  & 522.7189& 12560.930& -0.969 & 4.24  & 4.49&   -  &  -  & 14.43 & 5.14 & 19.24 & 5.73 & 11.86 & 5.01\cr 
Fe I  & 523.2939& 42815.858& -0.140 & 3.34  & 4.81& 5.67 & 5.42& 10.41 & 5.10 & 13.15 & 5.59 & 8.42  & 4.96\cr 
Fe II & 423.3172& 20830.582& -2.000 & 2.48  & 4.64& 4.39 & 4.92& 9.04  & 5.02 &  -    &  -   & 7.65  & 4.94\cr 
Fe II & 455.5893& 22810.357& -2.290 & -     &  -  & 3.58 & 5.25&   -   &  -   &  -    &  -   &  -    &  -  \cr 
Fe II & 492.3927& 23317.633& -1.320 & 4.32  & 4.57& 8.02 & 5.22& 10.84 & 4.95 & 12.80 & 5.57 & 9.19  & 4.83\cr 
Fe II & 501.8440& 23317.633& -1.220 & 5.12  & 4.60& 9.81 & 5.53& 12.45 & 5.14 & 14.83 & 5.72 & 10.03 & 4.89 \cr 
Fe II & 516.9033& 23317.633& -0.870 &  -    &  -  & 9.81 & 5.17&   -   &   -  &   -   & -    &  -    &   -  \cr 
Sr II & 407.7709& 0.000    & 0.167  & 6.90  & -0.40& 6.81& -0.33& 17.68& 0.25& 20.37 & 0.50 & 17.08 & 0.55\cr 
Sr II & 421.5519& 0.000    & -0.145 & 6.00  & -0.29& 6.90& -0.02& 14.71& 0.12& 19.47 & 0.70 & 14.74 & 0.51\cr 
Ba II & 455.4029& 0.000    & 0.170  &  < 1.3    & <-1.4  & 9.34& 0.48 & 13.16& -0.34& 11.08 & -0.67 & 11.08 & -0.34\cr 
\multispan{14}{\hrulefill}\cr 
} 
\endtable

\PBtbltwo{Errors on abundances for CS29529-12} 
\halign to \ColumnWidth{ 
\tabskip=15ptplus32ptminus15pt 
#& 
$#$\PBs& 
$#$\PBs& 
$#$\PBs& 
$#$\PBs& 
$#$ 
\cr 
\multispan{6}{\hrulefill}\cr 
Element & \sigma &\rm\Delta_{log~g+0.5} &\rm\Delta_{T-130 K} & \rm\Delta_{\xi+0.3 km/s}& \rm \Delta_{EW+0.4 pm} 
\cr 
\multispan{6}{\hrulefill}\cr\cr 
Na&  0.08 & -0.31 & -0.26 & -0.25 & +0.09 \cr 
Mg&  0.17 & -0.15 & -0.11 & -0.08 & +0.05 \cr 
Al&   -   & -0.01 & -0.11 & -0.15 & +0.09 \cr 
Ca&  0.15 & -0.11 & -0.17 & -0.09 & +0.08\cr 
Sc&  -    & +0.17 & -0.06 & -0.10 & +0.09 \cr 
Ti&  0.20 & +0.17 & -0.05 & -0.12 & +0.10\cr 
Cr&  0.16 & -0.01 & -0.12 & -0.06 & +0.08\cr 
Fe&  0.21 & +0.02 & -0.11 & -0.13 & +0.09 \cr 
Sr&   -   & +0.15 & -0.09 & -0.14 & +0.12\cr 
Ba&   -   & +0.12 & -0.08 & -0.27 & +0.11\cr 
\multispan{6}{\hrulefill}\cr 
} 
\endtable

\PBtbltwo{Abundances} 
\halign to \PageWidth{ 
\tabskip=15ptplus32ptminus15pt 
#& 
$#$\PBs& 
$#$\PBs& 
$#$\PBs& 
$#$\PBs& 
$#$\PBs& 
$#$\PBs& 
$#$\PBs& 
$#$\PBs& 
$#$\PBs& 
$#$\PBs& 
$#$ 
\cr 
\multispan{12}{\hrulefill}\cr 
 & \hfill\epsilon_\odot \hfill & \epsilon & {\rm [X/Fe]} 
& \hfill\epsilon\hfill & {\rm [X/Fe]} 
& \hfill\epsilon\hfill & {\rm [X/Fe]} 
& \hfill\epsilon\hfill & {\rm [X/Fe]} 
& \hfill\epsilon\hfill & {\rm [X/Fe]} 
\cr\cr 
 & & 
\multispan2{\hfill {\rm CS 22894-19}\hfill}& 
\multispan2{\hfill {\rm CS 29529-12}\hfill}& 
\multispan2{\hfill \rm BD $-14^\circ 5890$\hfill}& 
\multispan2{\hfill \rm HD 165195\hfill}& 
\multispan2{\hfill \rm HD 26169\hfill} 
\cr 
\multispan{12}{\hrulefill}\cr\cr 
Na&6.33& 3.13 &-0.29 & 5.62: &1.56: &  & &  & &  &\cr 
Mg&7.58& 5.04 &+0.36 & 5.85 &+0.54 & 5.61 &+0.55 & 5.74 &+0.08 & 5.46 &+0.49\cr 
Al&6.47& 3.01 &-0.56 & 3.70 &-0.50 & 3.14 &-0.81 & 3.90 &-0.65 & 3.16 &-0.72\cr 
Ca&6.36& 3.80 &+0.34 & 4.41 &+0.32 & 4.10 &+0.26 &  -   &  -   & 4.07 &+0.32\cr 
Sc&3.10& 0.07 &-0.13 & 0.74 &-0.09 & 0.42 &-0.16 & 1.12 &-0.06 & 0.44 &-0.07\cr 
Ti&4.99& 2.36 &+0.27 & 3.03 &+0.31 & 2.70 &+0.23 & 3.60 &+0.53 & 2.63 &+0.25\cr 
Cr&5.67& 2.64 &-0.13 & 3.43 &+0.03 & 2.90 &-0.25 & 3.50 &-0.25 & 2.72 &-0.36\cr 
Mn&5.39& 2.18 &-0.31 &   -  &  -   & 3.06 &+0.19 &  -   &  -   & 2.68 &-0.12\cr 
Fe&7.51& 4.61 & 0.00 & 5.24 & 0.00 & 4.99 & 0.00 & 5.59 & 0.00 & 4.90 & 0.00\cr 
Sr&2.90&-0.34 &-0.34 &-0.02 &-0.65 & 0.18 &-0.20 & 0.61 &-0.37 & 0.53 &+0.24\cr 
Ba&2.13&<-1.40 &< -0.60  &+0.48 &+0.62 &-0.34 &+0.05 &-0.67 &-0.88 &-0.34 &+0.13\cr 
\multispan{12}{\hrulefill}\cr 
\multispan{12}{a : denotes an uncertain value\hfill}\cr 
} 
\endtable 
 
\PBtbltwo{kinematical data} 
\halign to 
\PageWidth 
{ 
\tabskip=15ptplus32ptminus15pt 
#& 
\hfill $#$& 
\hfill $#$& 
\hfill $#$& 
\hfill $#$& 
 $#$& 
 $#$& 
 $#$ 
\cr 
\multispan{8}{\hrulefill}\cr 
Star  \hfill & \pi & \mu_\alpha  & \mu_\delta & v_r & U & V & W \cr 
& mas & mas  & mas  & \rm kms^{-1} &\rm kms^{-1} &\rm kms^{-1} &\rm kms^{-1} \cr 
\multispan{8}{\hrulefill}\cr 
BD -14$^\circ$ 5890 & 1.48\pm 1.86 & -16.96\pm 2.05 & -83.97 \pm 1.67 & +118 & -205 \pm 158 & -183 \pm 300 & -116 \pm 63\cr 
                    & 0.57^a & & & & -406 ~ \phantom{\pm 100} & -564 ~ \phantom{\pm 100} & -196 ~ \phantom{\pm 100}\cr 
HD 165195           & 2.20\pm 1.04 & -28.55\pm 0.83 & -82.87\pm 0.68 & 0  & -102\pm 48 & -157\pm 74 & -26 \pm 13\cr 
                    & 2.34^a & & & & -96 ~ \phantom{\pm 100} & -148 ~ \phantom{\pm 100} & -26 ~ \phantom{\pm 100}\cr 
HD 26169            & 2.83 \pm  0.79 & 137.18 \pm 0.80 & 74.90\pm 1.04 & -20 & 203 \pm 55 & -129 \pm 42 & 106 \pm 28\cr 
                    & 1.95^a & & & & 292 ~ \phantom{\pm 100} & -194 ~ \phantom{\pm 100} & 148 ~ \phantom{\pm 100}\cr 
 HD6446     &     1.96 \pm   0.57 &    38.98 \pm    0.53 &   -48.69 \pm    0.59 &   62 &    6\pm  10 & -153\pm  33 &   55\pm  28\cr 
 XX And     &     1.50 \pm   2.15 &    58.97 \pm   1.64 &   -35.46 \pm   1.47 &  -25 &  118\pm 189 & -170\pm 218 &  -72\pm 117\cr 
$^b$ & 1.07\phantom{\pm   2.15}& 55.76\phantom{\pm   1.64}& -35.16\phantom{\pm   1.47}&  44 &  118 
\phantom{\pm 189} & -170\phantom{\pm 218} &  -72\phantom{\pm 117}\cr 
 HD11569    &     9.47 \pm   0.77 &   268.97 \pm    0.71 &   -29.45 \pm   0.93 &  -18 &   97\pm   8 &  -80\pm  10 &   54\pm   8\cr 
 HD23592    &     1.67 \pm   1.00 &    14.37 \pm    0.98 &    25.68 \pm   1.06 &  -54 &   78\pm  50 &   36\pm   7 &   50\pm   9\cr 
 HD27928    &     2.36 \pm   1.16 &    26.90 \pm    0.90 &   -67.47 \pm   1.22 &   15 &  -98\pm  51 & -107\pm  48 &   22\pm  17\cr 
 HD33771    &     1.64 \pm   1.08 &    79.65 \pm    0.99 &   -30.25 \pm   1.11 &   18 &  -45\pm  35 & -179\pm 110 &  164\pm 115\cr 
 CD-33 3337 &     9.11 \pm   1.01 &  -178.43 \pm    0.90 &  -148.74 \pm   0.97 &   62 &  -15\pm   6 &  -44\pm   9 & -128\pm  13\cr 
 HD81223    &     3.39 \pm   0.69 &  -102.14 \pm    0.82 &    85.68 \pm    0.66 &  -52 &  193\pm  37 &    4\pm  13 &   -7\pm   5\cr 
 HD83212    &     1.96 \pm   0.98 &   -15.51 \pm    0.92 &   -20.41 \pm    0.65 &  110 &   26\pm   3 & -123\pm  16 &  -14\pm  29\cr 
 HD97320    &    17.77 \pm   0.76 &   159.19 \pm    0.74 &  -201.28 \pm    0.74 &   48 &  -72\pm   5 &  -18\pm   9 &  -38\pm   2\cr 
 BD-01 2582 &     2.98 \pm   1.35 &   -21.19 \pm   1.34 &  -117.77 \pm    0.78 &    0 &  -61\pm  28 & -151\pm  69 &  -99\pm  46\cr 
 HD111721   &     3.29 \pm   1.11 &  -273.70 \pm    0.94 &  -321.99 \pm    0.66 &   24 &  131\pm  47 & -524\pm 173 & -282\pm 101\cr 
 HD118055   &      .66 \pm   1.24 &   -23.09 \pm   1.64 &   -14.69 \pm    0.94 & -101 &  145\pm 172 & -121\pm 317 & -115\pm  82\cr 
 HD122196   &     9.77 \pm   1.32 &  -452.86 \pm   1.08 &   -82.55 \pm   1.01 &  -24 &  172\pm  22 & -144\pm  22 &   16\pm   5\cr 
 HD136316   &     1.08 \pm   0.95 &   -27.55 \pm    0.86 &   -64.44 \pm    0.83 &  -45 &  178\pm 125 & -189\pm 189 & -171\pm 149\cr 
 EZ Lyr$^b$ &     0.95\phantom{\pm   0.95} &-1.59\phantom{\pm    0.86} & 13.10\phantom{\pm    0.83}&-60.00& 
95\phantom{\pm 125}&-33\phantom{\pm 189}&24\phantom{\pm 149}\cr 
 SW Aqr     &     1.15 \pm   2.46 &   -42.05 \pm    2.63 &   -58.87 \pm   1.96 &  -62 &  -207\pm 511 & -219\pm 381 &   44\pm  27\cr 
$^b$ & 0.74 \phantom{\pm   1.01} & -42.89 \phantom{\pm    2.63} & -59.11 \phantom{\pm    1.96} & -36 &  -274 
\phantom{\pm   511} &   -240\phantom{\pm   381} & 36\phantom{\pm  27}\cr 
G090-003 & 1.20\pm 1.41 & 16.93 \pm 2.36 & -199.64\pm  2.44& 29 & 14\pm 26 & -774\pm 1516 & -168 \pm 352\cr 
   BD+80245 &     3.91 \pm   0.88 &   136.81 \pm   1.22 &  -366.66 \pm   0.99 &    4 &  183\pm  56 & -366\pm 115 &  241\pm  75\cr 
G182-031  &     5.91 \pm   0.98 &  -154.86 \pm   1.17 &  -244.96 \pm   1.16 &  -61 & -158\pm  36 & -178\pm  27 &   32\pm  13\cr 
G023-014  &     8.11 \pm   4.40 &  -140.42 \pm   6.21 &  -231.74 \pm   7.71 &   18 & -126\pm  87 &  -98\pm  85 &    1\pm   6\cr 
G190-015  &    14.09 \pm   1.27 &   175.38 \pm   2.20 &  -313.55 \pm   1.86 &  -55 &   -8\pm   2 &  -93\pm  11 &  -95\pm  18\cr 
\multispan{8}{\hrulefill}\cr 
\cr 
\multispan{8}{$^a$ photometric, estimated from $M_v$ derived from isochrones 
\hfill}\cr 
\multispan{8}{$^b$ distance from Layden (1994), proper motions from Martin\&Morrison (1998) 
\hfill}\cr 
} 
\endtable

\section*{Acknowledgments} 
We are grateful to J. Danziger for critically reading
our manuscript and suggesting many improvements.
Special thanks are due to T. Valente for his help in the measuring 
of equivalent widths.
Use was made of the SIMBAD data base, operated at the CDS,
Strasbourg, France. 
This research was partially supported by Collaborative NATO grant No. 950875. 
This research is based on observations collected at the European Southern 
Observatory, Chile.

 \appendix
\section{Comparison with previous results} 
 
For BD-14$^\circ 5890$, the only high-resolution studies  
which allow a comparison to are those of Pilachowski, Sneden \& Booth (1993) 
and 
Cavallo, Pilachowski \& Rebolo (1997). 
The [Fe/H] and [Ca/H] of Pilachowski \PBal (1993) are 0.45 dex  
and 0.54 dex respectively 
higher than ours. 
This discrepancy can  
be largely attributed to the difference  
in the adopted effective temperatures:  
our \PBt ~ is 
250 K cooler.  
The  [Fe/H] of  Cavallo \PBal  
is 0.26 dex higher than ours and in this case the main 
cause of the discrepancy is the different microturbulent 
velocity.  
However, if we adopt the value of  
Cavallo \PBal (0.8 \PBkms ) our  
[Fe/H] becomes 0.4 dex {\it larger} than that 
of Cavallo \PBal; incidentally our line to line 
scatter doubles with this low microturbulent velocity.

For HD 165195 we compared our results 
with those of Francois (1996), 
Gratton \& Sneden (1991) and Gilroy \PBal (1988). 
With respect to Francois our [Fe/H] is 0.25 dex larger, 
and this may be attributed 
to our lower microturbulent velocity. In 
contrast there is a  
strong disagreement in the Ba abundance, ours is 
about 0.9 dex {\it lower} than that of Francois. We have 
no explanation for this difference, but point out 
that the result of Francois was based only on the 614.17 nm line 
while  ours only on the 455.4 nm line. 
With Gratton \& Sneden the agreement is generally good 
at the 0.1 - 0.2 dex level, with the exception 
of Ti where our abundance is over 0.3 dex larger. 
Also in this case our  lower microturbulent velocity 
is the main responsible for the discrepancy. 
Also the comparison with Gilroy \PBal   
is  generally good, again with the notable 
exception of Ti (ours is 0.66 dex larger), 
The microturbulence adopted by Gilroy \PBal is the 
highest for this star 
among the reviewed studies: 2.8 \PBkms. With this 
high value of $\xi$ our data imply a Ti abundance which is 
0.73 dex lower, i.e. in substantial agreement 
with that of Gilroy \PBal. 
 
For HD 26169 we compared our results with those of 
Peterson, Kurucz \& Carney (1990);  
for all the elements in common our abundances are 
lower than those of Peterson \PBal,  
 only a few hundredths of dex for Sc and Ba and as much as 0.36 
dex for Cr.  
While this agreement may be regarded as satisfactory in the context 
of the claimed accuracy of the present analysis, it is 
somewhat disappointing, considering the fact that we 
have virtually the same atmospheric parameters  
(the only difference is 0.1 \PBkms in $\xi$) and adopt 
the same kind of models.

\bye